# Generic-Precision algorithm for DCT-Cordic architectures

Imen Ben Saad[1], Younes Lahbib[3], Yassine Hachaïchi[2,3], Sonia Mami[1]

and Abdelkader Mami[1]

[1]Université de Tunis El Manar,
Faculté des Sciences de Tunis, LACS-ENIT
`soniammkd@gmail.com`
`bensaad_imen1@yahoo.fr`
`abdelkader.mami@fst.rnu.tn`
[2]ENIT-LAMSIN
Université Tunis Al Manar
`Yassine.Hachaichi@ipeiem.rnu.tn`
[3]ENICarthage
Université de Carthage
`Younes.Lahbib@enicarthage.rnu.tn`

**Abstract.** In this paper we propose a generic algorithm to calculate the rotation parameters of CORDIC angles required for the Discrete Cosine Transform algorithm (DCT). This leads us to increase the precision of calculation meeting any accuracy. Our contribution is to use this decomposition in CORDIC based DCT which is appropriate for domains which require high quality and top precision.

We then propose a hardware implementation of the novel transformation, and as expected, a substantial improvement in PSNR quality is found.

## 1 Introduction

The Discrete Cosine Transform DCT was developed by Ahmed et.al in 1974[1]. It is a robust approximation of the optimal Karhunen-Loeve Transform (KLT) [2]. It has become one of the most widely used techniques of transforms in digital signal processing. The DCT is one of the computationally intensive transforms. It requires many multiplication and additions. Many researches had been done on low-power DCT designs [3, 4]. In consideration of VLSI-implementation, Flow-Graph Algorithm (FGA) is the most popular way to realize the fast DCT (FDCT) [5, 6].

Most existing works handled the reduction of energy consumption of the DCT. As the multiplications are energy expensive operations, several algorithms are based on additions and shifts instead of multiplications.

In 2004, Jeong et .al [4] suggested improving a Cordic based implementation of the DCT. COordinate Rotation Digital Computer (CORDIC) is an algorithm which can be used to evaluate various functions in signal processing [7, 8]. In [4], authors proposed a low-complexity CORDIC based DCT algorithm based on FGA.It requires only 38 add and 16 shift operations and consumes about26.1 % less power compared to [9],with a minor image quality degradation of 0.04 dB.

In the same direction, Sun et .al [10, 11] proposed a new flow graph for Cordic based Loeffler DCT implementation. A new table of parameters is obtained with new choice of the elementary rotations. However, Authors did not give any demonstrations or a method to compute the underlying elementary rotations.Their experimental result shows that the Cordic-based Loeffler DCT consumes 16%of energy compared to [15] with a minor image quality degradation of 0.03 dB.

After this analysis of state of the art, we remark that previous works chose to reduce precision of calculation to decrease the energy consumption.Themain disadvantage of these methods is the lack of precision, which restricted their applications in Medical, Biometrics and High frequency for example.

Despite of 0.01w power loss, our proposed DCT architecture reaches an improved image quality of 6.55 dB PSNR compared to [10, 11].This gain is very significant compared to PSNR evaluation of previous works, (see TABLE III in [13]). For thisreason, we proposeincreasing the precision degree ofDCT-Cordic micro-rotations with an exact and generic method.According to a chosen precision degree, exact DCT parameters are computed using the proposed algorithm.

This paper is organized as follows. Section 2 briefly introduces the algorithms of conventional Cordic-Based DCT Architecture. In Section 3, we present the solution approach of the high precision Cordic-Rotation. The experimental results are shown in Section 4 while Section 5 concludes this paper.

## 2  Conventional Cordic-Based DCT Architecture

### 2.1  Cordic Algorithm

The conventional Cordic algorithm [7] is hardware-efficient used for the approximation computation of the transcendental functions. It only uses shift and addition operations. The Cordic algorithm can operate in two modes, namely vectoring and rotation. In this paper we focus on the first mode.

In the conventional Cordic algorithm, a rotation angle is decomposedinto combination of micro-rotation angles of arctangent radix. When the vector is rotated by an angle θ, the coordinate changed from $(X_i, Y_i)$ to $(X_{i+1}, Y_{i+1})$.

The value of vector after this micro rotation can be represented as:

$$\begin{bmatrix} x_{i+1} \\ y_{i+1} \end{bmatrix} = K_i \begin{bmatrix} 1 & -\sigma_i \tan \theta_i \\ \sigma_i \tan \theta_i & 1 \end{bmatrix} \begin{bmatrix} x_i \\ y_i \end{bmatrix} \qquad (1)$$

where $K_i = \cos \theta_i$ and $\theta_i = \tan^{-1}(2^{-i})$

The circular rotation angle is depicted as:

$$\theta = \sum_i \sigma_i . \theta_i \quad , \text{where } \sigma_i = \pm 1 \qquad (2)$$

Then, the vector rotation can be performed iteratively as follows:

$$\begin{bmatrix} x_{i+1} \\ y_{i+1} \end{bmatrix} = K_i \begin{bmatrix} 1 & -\sigma_i 2^{-i} \\ \sigma_i 2^{-i} & 1 \end{bmatrix} \begin{bmatrix} x_i \\ y_i \end{bmatrix} \qquad (3)$$

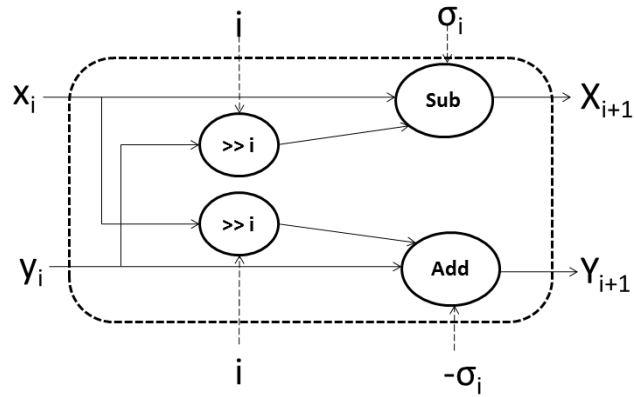

**Fig. 1.** The direct implementation of equation 3.

In the equation 3, only shift and add operations are required to perform the rotation angle described in Fig. 1. But, the results of the rotation iterations need to be scaled by a compensation factor K. This can be done by using the following iterative method.

$$K = \prod_i k_i = \prod_i 1/\sqrt{1 + 2^{-2i}} \qquad (4)$$

The scale factor K which can be interpreted as a constant gain (hence not data dependent) can be tolerated in many digital signal processing applications. Hence, it should be carefully investigated whether it is necessary to compensate for the scaling at all. If scale factor correction cannot be avoided, two possibilities are known. The first approach consists on performing a constant factor multiplication with $1/K_i$. The second method is based on extending the Cordic iteration in a way that the resulting inverse of the scale factor takes a value. In other words, we need to write the scaling factor as a sum of $2^{-i}$ where i must be determined so that the error is minimized. In the rotation mode, the angle accumulator is initialized with the desired rotation angle $\theta$. The rotation decision at each iteration is made to diminish the magnitude of the resi-dual angle in the accumulator one. The decision at each iteration is therefore based on the sign of the residual angle after each step [12].

## 2.2 Cordic-Based DCT Architecture

The One-dimensional DCT for 8x8 sub-images is defined as

$$F(k)=\frac{1}{2} C(k) \sum_{x=0}^{7} f(x)\cos[\frac{(2x+1)k\pi}{16}]$$
$$C(k)=\begin{cases} \frac{1}{\sqrt{2}} & if\ k = 0 \\ 1 & otherwise \end{cases} \quad (5)$$

*Where x (i) is the input data and X (t) is 1-D DCT transformed output data.*
The two-dimensional DCT is a separable transform. It can be executed by one-dimensional DCT in a serial manner as shown in the Fig. 2.

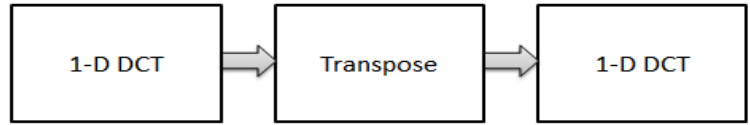

**Fig.2**. 8 x 8 2-D DCT processor with separable 1-D DCT

The 1-D DCT transform is represented as follows.

$$\begin{bmatrix} x_0 \\ x_2 \\ x_4 \\ x_6 \end{bmatrix} = \frac{1}{2}\begin{pmatrix} A & A & A & A \\ B & C & -C & -B \\ A & -A & -A & A \\ C & -B & B & -C \end{pmatrix}\begin{bmatrix} x_0 + x_7 \\ x_1 + x_6 \\ x_2 + x_5 \\ x_3 + x_4 \end{bmatrix}$$
$$\begin{bmatrix} x_1 \\ x_3 \\ x_5 \\ x_7 \end{bmatrix} = \frac{1}{2}\begin{pmatrix} D & E & F & G \\ E & -G & -D & -F \\ F & -D & G & E \\ G & -F & E & -D \end{pmatrix}\begin{bmatrix} x_0 - x_7 \\ x_1 - x_6 \\ x_2 - x_5 \\ x_3 - x_4 \end{bmatrix} \quad (6)$$

where $A=\cos\left(\frac{\pi}{4}\right)$, $B=\sin\left(\frac{3\pi}{8}\right)$, $C=\cos\left(\frac{3\pi}{8}\right)$, $D=\sin\left(\frac{7\pi}{16}\right)$, $E=\cos\left(\frac{3\pi}{16}\right)$, $F=\sin\left(\frac{3\pi}{16}\right)$ and $G=\cos\left(\frac{7\pi}{16}\right)$.

The rearranged 1-D DCT equation is now represented as a vector rotation matrix together with the consecutive CORDIC iterations as shown in Fig.3. The Cordic array performs the fixed-angle rotation in the DCT algorithm. Therefore, the general signal flow graph of Cordic-based DCT is presented by Fig. 4.

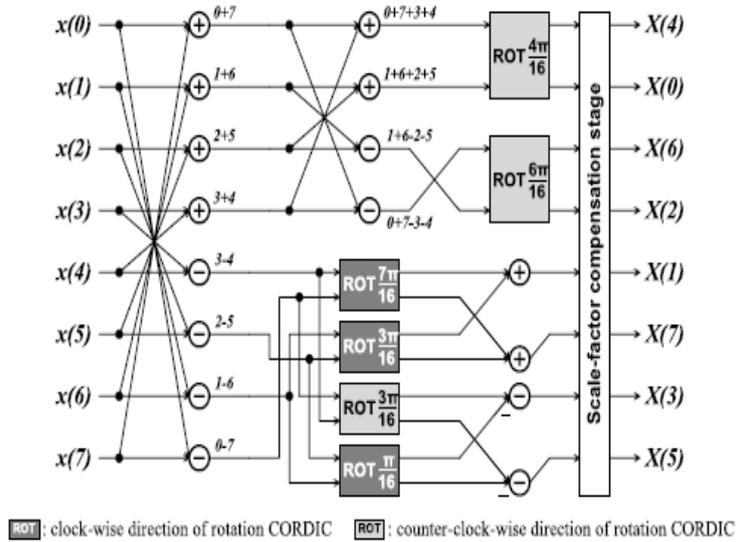

**Fig. 3.** Hardware architecture of CORDIC-based 1-D DCT[13]

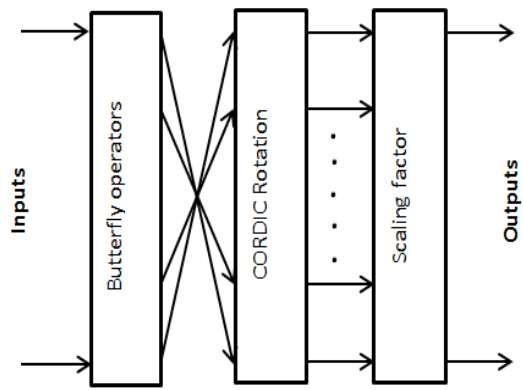

**Fig. 4.** The general signal flow graph CORDIC-based DCT

According to the Fig. 4, the signal flow can be represented by three major components, the butterfly operator, the fixed-angle CORDICs and the post-scaling factors of 8-point DCT.

# 3 Proposed High precision Cordic-Based Loeffler DCT Architecture

In this section we will present the proposed algorithm which calculates the Cordic Rotations. The main result of this algorithm is enhancing the degree of precision by improving the selected parameters in order to find the exact values of the rotations.

## 3.1 Computation of Micro-Rotation decomposition

The proposed algorithm takes as input the rotation angle and the precision degree. The choice of this last parameter will depend on the application field. For instance, medical or biometrics applications require high precisions whereas compression process or video surveillance can be treated with low a greater tolerance error.

| *Algorithm obtains the optimal parameter of micro-rotations.* |
|---|
| Input : $\theta = angle$ and $\varepsilon = precision$ <br> 1: while $|\theta| > \varepsilon$ do <br> 2: $x = \tan|\theta|$ <br> 3: $i = \lfloor -\log(x) \rfloor + 1$ <br> 4: if $\theta > \tan^{-1}(2^{-i})$ then <br> 5: $\sigma = +$ <br> 6: $\theta = \theta - \tan^{-1}(2^{-i})$ <br> 7: else <br> 8: $\sigma = -$ <br> 9: $\theta = \theta + \tan^{-1}(2^{-i})$ <br> 10: write $i$ and $\sigma$ <br> 11: end while |

This algorithm provides the Cordic parameters (iterations and direction) corresponding to the angle and the selected precision. The iterations, in other words, the micro-rotations are identified by the algorithm with their orientation, clockwise or anticlockwise.

When applying our algorithm we obtain the new parameters table which depends on the required precision. (See Table 1)

**Table 1.** Required Iterations and Directions for Vector Rotation

| Angle / Precision | $\theta = \dfrac{\pi}{4}$ | $\theta = \dfrac{3\pi}{8}$ | $\theta = \dfrac{\pi}{16}$ | $\theta = \dfrac{3\pi}{16}$ |
|---|---|---|---|---|
| $P=10^{-3}$ | i=0<br>$\sigma_i$= - | i=0/1/4/7<br>$\sigma_i$=- + - - | i=2/4/6/9<br>$\sigma_i$=+ - + - | i= 1/3/10<br>$\sigma_i$=+ + + |
| $P=10^{-4}$ | i=0<br>$\sigma_i$= - | i=0/1/4/7/10/12<br>$\sigma_i$=+ + - - - + | i=2/4/6/9/13<br>$\sigma_i$=+ - + - + | i= 1/3/10<br>$\sigma_i$=+ + + |

Let's take an example and show its implementation. (See Table 2)

**Table 2.** Rotator parameter of the $\dfrac{\pi}{16}$ angle for precision $P = 10^{-4}$

| Input | | Output | | | | | |
|---|---|---|---|---|---|---|---|
| $\theta$ | $P$ | $i$ | 2 | 4 | 6 | 9 | 13 |
| $\dfrac{\pi}{16}$ | $10^{-4}$ | $\sigma_i$ | + | - | + | - | + |

So we can write the rotation angle as the weighted sum of micro-rotations $i$ as shown in the equation 2 and 7, we obtain:

$$\theta = \frac{\pi}{16} = \theta_2 - \theta_4 + \theta_6 - \theta_9 + \theta_{13} = 0.196349$$

So for the precision $P = 10^{-4}$ and since the real value of $\theta = 0.1963$ we can conclude that we have a zero-error approximation in the angle. Based on the previous computed micro-rotations of the $\dfrac{\pi}{16}$ angle, the Cordic architecture computing $\dfrac{\pi}{16}$ angle is given in Fig. 5.

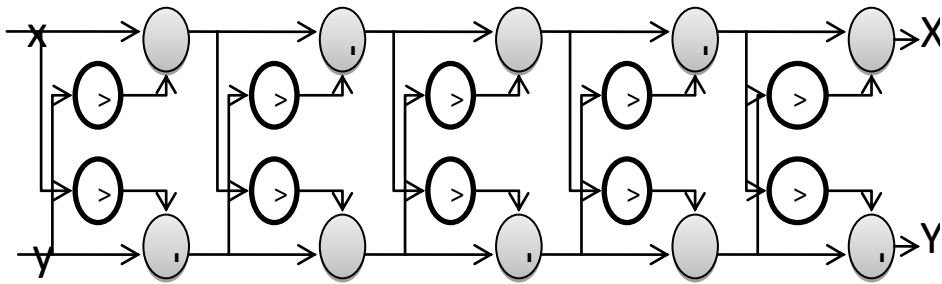

**Fig. 5.** Unfolded flow graph of the π/16 angle

In order to check the correctness of the computed rotations matrices according to the proposed algorithm, obtained values are compared to ideal ones. Like seen in the equation 3 and since we have 5 iterations, the Cordic of the angle $\theta = \frac{\pi}{16}$ is obtained by multiplying 5 rotation matrices:

$$\begin{bmatrix}X\\Y\end{bmatrix} = K \begin{bmatrix}1 & -2^{-2}\\2^{-2} & 1\end{bmatrix}\begin{bmatrix}1 & 2^{-4}\\-2^{-4} & 1\end{bmatrix}\begin{bmatrix}1 & -2^{-6}\\2^{-6} & 1\end{bmatrix}\begin{bmatrix}1 & 2^{-9}\\-2^{-9} & 1\end{bmatrix}\begin{bmatrix}1 & -2^{-13}\\2^{-13} & 1\end{bmatrix}\begin{bmatrix}x\\y\end{bmatrix} \quad (7)$$

Then we get the following matrix coefficients according to the proposed algorithm (equation 8).

$$\begin{bmatrix}X\\Y\end{bmatrix} = K\begin{bmatrix}1.013067933963612 & -0.2015148886130191\\0.2015148886130191 & 1.013067933963612\end{bmatrix}\begin{bmatrix}x\\y\end{bmatrix} \quad (8)$$

The ideal computed coefficients of $\frac{\pi}{16}$ rotation matrix are depicted in equation 9.

$$\begin{bmatrix}X\\Y\end{bmatrix} = K\begin{bmatrix}1.014359843600569 & -0.201768717865449\\0.201768717865449 & 1.014359843600569\end{bmatrix}\begin{bmatrix}x\\y\end{bmatrix} \quad (9)$$

As it can be denoted, the ideal rotation matrix (equation 8) has exactly the same coefficients of the computed rotation matrix (equation 9), for an approximation of $10^{-3}$

## 4  Experimental Results

Our algorithm has been implemented on Virtex5 xc5vlx30-3ff676 using Xilinx System Generator (XSG). The power consumption of our DCT architecture with different precision degrees is shown in the Table 3. The power consumption is measured with Xpower Analyzer with 100 Mhz clock cycles, 1V supply power.

**Table 3.** Hardware Implementation and PSNR from high-to-low compression quality in JPEG2000 for Lena image

|  | Power (W) | Quality factor | PSNR |
|---|---|---|---|
| **[10, 11]** | 0.642 | Q=95 | 36.982 |
|  |  | Q=90 | 36.019 |
|  |  | Q=85 | 35.113 |
|  |  | Q=80 | 34.295 |
|  |  | Q=75 | 33.708 |
| **Proposed algorithm (P=$10^{-3}$)** | 0.656 | Q=95 | 43.535 |
|  |  | Q=90 | 39.458 |
|  |  | Q=85 | 36.992 |
|  |  | Q=80 | 35.355 |
|  |  | Q=75 | 34.280 |
| **Proposed algorithm (P=$10^{-4}$)** | 0.659 | Q=95 | 43.538 |
|  |  | Q=90 | 39.459 |
|  |  | Q=85 | 36.994 |
|  |  | Q=80 | 35.355 |
|  |  | Q=75 | 34.281 |

In order to demonstrate the high-quality feature of the proposed DCT algorithm, it has been evaluated considering a JPEG2000 compression chain [14].

Table 3 demonstrates the comparison of the average PSNR of the proposed DCT algorithm in tow precision degrees, with the CORDIC based Loeffler DCT [10, 11]. Checked results considerhigh-to-low quality compression (i.e. quantization factors from 95 to 75) using some well-known test images. For instance, Fig. 8 gives the experimental results based on Lena image.

It can be easily noticed from Table 2 that the performance of the proposed DCT algorithm has better quality about 6.55 for Q=95 dB than the Cordic-based Loeffler for the JPEG2000 standard, without substantial loss of power.

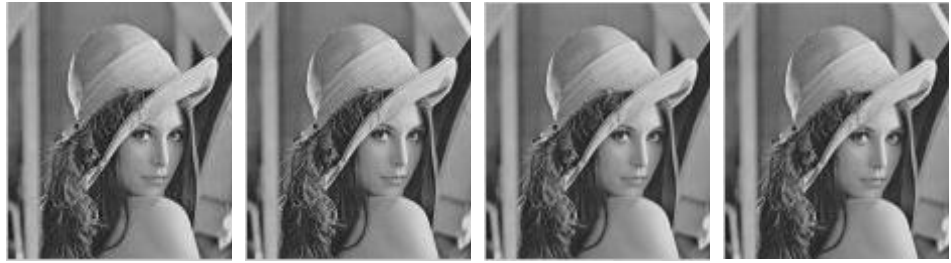

(a) 34.280 dB for Q = 75    (b) 36.992 dB for Q = 85    (c) 43.535 dB for Q = 95    (d) 36.982 dB for Q = 95

**Fig. 8.** Lena images obtained using the proposed Cordic-based Loeffler DCT for $P = 10^{-3}$. (a) Q=75. (b) Q=85. (c) Q=95. (d) Lena image obtained using [10, 11]

## 5 Conclusion

In this paper, we proposed a high quality algorithm which calculates the rotation parameters of CORDIC angles required for the Cordic-based Loeffler DCT algorithm. The obtained results show a significant improvement in the PSNR (6, 55 dB for Q=95) without a substantial loss of Power. These enhancements make our algorithm appropriate for high accuracy applications such as medical or biometrics domains.

The proposed algorithm calculates the rotation parameters of CORDIC for different precision degrees from $10^{-3}$ up to $10^{-7}$. The precision degree results higher than $10^{-4}$ they don't give a great improvement in terms of PSNR with respect to the precision $10^{-3}$. In fact, for Q=95, the PSNR obtained for a precision of $10^{-6}$ is equal to 43.53831 dB, which means that the variation is in the order of $10^{-4}$ dB. This improvement is small especially that it requires the addition of some layers of rotations in the CORDIC architectures. Hence, it is useless to go higher than a precision of $10^{-3}$.